\documentclass[twocolumn]{revtex4}

\def\be{\begin{equation}}
\def\ee{\end{equation}}
\def\la{\label}
\def\bea{\begin{eqnarray}}
\def\eea{\end{eqnarray}}
\def\non{\nonumber}
\def\ci{\cite}
\def\la{\label}
\def\bib{\bibitem}

\def\le{\left}
\def\ri{\right}

\def\s8{\sigma_8}

\def\fr{\frac}
\def\pp{\partial}

\def\non{\nonumber}

\def\rp{\rho_\phi}
\def\rpo{\rho_{\phi o}}
\def\wp{w_\phi}

\def\Ompo{\Omega_{\phi o}}
\def\wpo{w_{\phi o}}
\def\wpe{w_{\phi eff}}

\def\Omp{\Omega_\phi}

\def\we{w_{ eff}}
\def\Omp{\Omega_\phi}

\def\r{\rho}

\def\rp{\rho_\phi}
\def\rpi{\rho_{\phi i}}
\def\rpo{\rho_{\phi o}}

\def\rb{\rho_\psi}
\def\rbi{\rho_{\psi o}}
\def\rbo{\rho_{\psi o}}

\def\rf{\rho_\psi}

\def\rfo{\rho_{\psi o}}
\def\wf{w_\psi}

\def\wfe{w_{\psi eff}}
\def\Of{\Omega_\psi}
\def\Ofo{\Omega_{\psi o}}

\def\wp{w_\phi}

\def\Ompo{\Omega_{\phi o}}
\def\wpo{w_{\phi o}}
\def\we{w_{eff}}

\def\wa{w_{app}}

\def\G{\Gamma}

\usepackage{graphicx}

\begin{document}

\title{Interacting Dark Energy: Decay into Fermions }

\author{A. de la Macorra }
\affiliation{Instituto de F\'{\i}sica UNAM\\ Apdo. Postal 20-364\\
01000 M\'exico D.F., M\'exico}

\begin{abstract}

A dark energy component is responsible for the present stage
of acceleration of our universe. If no fine tuning is assumed
on the dark energy potential  then
it will end up dominating the universe at late times
and the universe will not stop this stage of acceleration.
On the other hand, the equation of state of dark energy
seems to be smaller than -1 as suggested by the cosmological data.
We take this as an indication that dark energy does indeed interact
with another fluid (we consider  fermion fields ) and
we determine the interaction through
the cosmological data  and extrapolate  it into the future.
We study the conditions under
which a dark energy can dilute faster or   decay into
the  fermion fields.
We show that it is   possible to
live now in an accelerating epoch dominated by the dark
energy and  without
introducing any fine tuning parameters
the dark energy can
 either dilute faster or decaying into fermions
 in the future.
The acceleration
of the universe will then cease.

\end{abstract}


\maketitle

\section{Introduction}

A dark energy component is probably responsible for the present stage
of acceleration of our universe\ci{DE},\ci{SN}.
Perhaps the most appealing candidate for  dark
energy is that of a scalar field, quintessence \ci{Q}, which can be
either a fundamental particle or a composite particle \ci{Qax}.
Since  dark energy dilutes slower than matter we expect it to
dominate the universe at late times if no fine tuning is assumed
on the dark energy potential $V$  and the universe will not stop the
present  stage of acceleration. However,
this conclusion can be overcome by assuming a
dark energy interaction \ci{fate}.

On the other hand,  the equation of state of dark energy
seems to be smaller than minus one as suggested by the cosmological data
\ci{DE},\ci{SN}. In general fluids with $w<-1$ give
many theoretically problems such as stability issues or wrong
kinetic terms as   phantom fields \ci{ph.etc}. However,  interacting
dark energy \ci{IDE}-\ci{IDE-ax},\ci{fate} is a very
simple and attractive  option which we will use in this {\it letter}.
The  interaction must be quite weak  since dark
 energy particles have not been produced in the accelerator
and because the dark energy has not decayed into lighter (e.g. massless) fields
such as the photon.

It was suggested
to determine  the interaction between dark energy and this
other fluid through the cosmological observations
and extrapolate it into the future. Doing so, it was shown within
a general framework  that the dark energy
dilutes exponentially fast  and  matter prevails in the end \ci{fate}.

In this letter we take the dark energy to interact with fermion
fields. It is tempting to take the  fermions as neutrinos \ci{IDE-n} since
the order of magnitude of the energy  of dark energy and
neutrinos is similar. Furthermore, if the mass of neutrinos
is larger than $0.8 eV$, as implied by the Heidelberg-Moscow
experiment,  then the dark energy cannot be a cosmological
constant \ci{IDE-ax}. However, in this {\it letter} we do not
assume  the fermions to be neutrinos.

We take the interaction between dark energy and fermions  through
a field dependent mass  $M=f(\phi)$, where $M$ is the mass of
the fermions \ci{IDE-n}.
The function $f(\phi)$ plays an important roll in the
evolution of the dark energy. The total potential $V_T$ of the scalar
field   has a contribution of the original
scalar potential $V(\phi)$ and the interaction term  $f(\phi)$.
We calculate the conditions for acceleration and we show  that
the naive slow roll conditions $V_T'/V_T\ll 1$ and $V_T''/V\ll 1$
{\it do not} imply   acceleration.
We determine the requirements  under
which a dark energy can dilute faster or decay into the fermion fields.
We obtain  that only in the
case where the mass of the dark energy, given by
$m^2=  V''+\rf f''/f$, is dominated by the interaction
term can the dark energy decay. We also show that
it is possible to have dark energy   diluting  faster than the
fermion fields.

This {\it letter} is organized as follows. In section \ref{de}
we present our assumptions about dark energy in the absence
of interactions while in section \ref{ide}
we present the dynamical equations  of dark energy
coupled to fermions  via a fields dependent mass.
In section \ref{DeWi} we present the generic requirements for
a particle decay and in section \ref{GeCo} we set the conditions
for dark energy to decay in the future. In section \ref{FlEv}
we briefly show the generic evolution of two coupled
fluids and we discuss the evidence for dark energy
decay. Finally, in section \ref{dedis} we study the possibility
of dark energy diluting or decaying into fermions and
we then present our conclusions in \ref{conc}.

\section{Dark Energy}\la{de}

We will consider in this {\it letter}
that the dark energy is given in terms of a scalar field with
lagrangian
\be\la{Lp}
  L(\phi)=\fr{1}{2}\;\dot\phi^2-V(\phi),
\ee
where $V(\phi)$ is only a function of $\phi$.
We define the energy density and pressure of
the scalar field as
\be\la{rp}
\rp\equiv \fr{1}{2}\dot\phi^2+V(\phi),\hspace{1cm}
p_\phi\equiv \fr{1}{2}\dot\phi^2-V(\phi),
\ee
i.e. without any interaction with other fields,
and
\be\la{wp}
\wp\equiv \fr{p_\phi}{\rp}=\fr{\dot\phi^2/2-V(\phi)}{\dot\phi^2/2+V(\phi)}.
\ee
Clearly $\wp\geq -1$ at all times if $\rp\geq 0$.
For $V$ to inflate
the slow roll constrains to the potential must
be satisfied,
\be\la{sr}
|V'/V| \ll 1, \hspace{1cm} |V''/V| \ll 1,
\ee
where a prime denotes derivative w.r.t. $\phi$, i.e. $V'\equiv dV/d\phi$.
We do not want to introduce any arbitrary
dark energy scale so we will assume that the
slow roll conditions are satisfied also in the future
and therefore   $V$ is a runaway potential
and tends to zero at $\phi\rightarrow \infty$.
From eqs.(\ref{sr}) we see that $V'$ and $V''$ also
approach zero at late times and
$V'$ is therefore negative. Inflation
occurs in general for $\phi \geq 1$ and the mass $m \simeq H$.

\section{Interacting Dark Energy}\la{ide}

The interaction term between two types of
particles  have two important contributions
to the relative ratio of energy densities
of these particles. The first one  has a classical
interpretation and it is due to
the cosmological evolution of the different
energy densities. This evolution will be
determined  the dynamics which includes the potential $V$
and the interaction term $L_{int}$ and is calculated by solving
the equation of motion of the lagrangian in a FRW
metric. The second consequence of the interaction
term $L_{int}$ is to allow for a particle decay,
which is due to quantum physics.
Both effects are relevant for determining the
ratio of energy densities.

\subsection{Interacting  Dark Energy with Fermions}\la{fer}

The interaction between  dark energy and fermions $\psi$ can be
achieved by taking a Yukawa type interaction \ci{IDE-n}
\be\la{int}
{\it L_{Int}}=
- f(\phi)  \overline{\psi}\psi
\ee
with a function $f$ that depends on the scalar field $\phi$. This
term gives an effective field dependent  mass   to $\psi$
\be\la{M}
M=f(\phi).
\ee
In principle the function $f$ is an arbitrary semipositive definite
function of $\phi$,
$f(\phi)\geq 0$. However, it is useful to
assume that it is a monotonic function, otherwise
not much can be said about $f$.

The evolution of the energy density of the fermions $\rf$ can be
determined using the  Fermi-Dirac distribution and taking into
account that the mass $M$ is a function of the scalar field
$\phi$,  the evolution of $\psi$ and $\phi$ are given by \ci{IDE-n}
\bea\la{rf}
\dot\rf+3H\rf(1+\wf) &=&   (1-3\wf)\rf \fr{\dot\phi f'}{f}\\
\ddot\phi+3H\dot\phi+V'&=&-(1-3\wf) \rf\fr{  f'}{f}
\la{rp0}
\eea
with $V(\phi)$ the potential for the scalar field
$\phi$ in the absence of interaction and $f'=df/d\phi$.
The solution to eq.(\ref{rf}) is
\bea\la{srf}
\rf &=& \rfo
a^{-3(1+\wf)} \le(\fr{f}{f_o}\ri)^{1-3\wf}
=\widetilde{\rf}\le(\fr{f}{f_o}\ri)^{1-3\wf}\non\\
\widetilde{\rf}&\equiv & \rfo a^{-3(1+\wf)}
\eea
where from now on the subscript $o$ is present time.
In this case the fermion
fluid no longer redshifts as $a^{-3(1+\wf)}$ since the evolution
of $f(\phi)$ also contributes to the redshift. Eq.(\ref{srf})
reduces to $ \rf=\rfo \;a^{-3  }\;  f /f_o$ when $\wf=0$ which was
used in \ci{IDE-n}. The quantity $\widetilde{\rf}$ is independent on
$\phi$. We see that for radiation $\wf=1/3$ the interaction term
is $\delta \equiv(1-3\wf)\rf \dot\phi f'/f=0$ and
$\rf=\widetilde{\rf}\sim a^{-4}$, consistent
with having a fermion field with a constant (vanishing) mass. From
now on we will set $\wf=0$  for simplicity   but the generalization
is straight forward.

Eqs.(\ref{rf})  and  (\ref{rp0}) can we rewritten in terms
of $\rp$, defined in eq.(\ref{rp}), and $\rf$ as
\bea\la{rmp}
\dot\rf+3H\rf(1+\wf) &=& \rf \fr{\dot f}{f}=\delta(t),  \\
\dot\rp+3H\rp(1+\wp) &=&  -\rf  \fr{\dot f}{f}=-\delta(t)
 \la{rmf}\eea
 where we have used $\dot f=f'\dot\phi$.
The interaction term  is defined by
$\delta\equiv \rf   \dot f/f$ is a function of time
and we could solve eqs.(\ref{rmp})
without making  reference to the scalar field $\phi$.
We can write
eqs.(\ref{rmp}) and (\ref{rmf})  as
\bea\la{rw}
\dot\rp &=& -3H\rp(1+\wpe )\non\\
\dot\rb &=& -3H\rb(1+\wfe)
\eea
with the effective equation states given by
\bea\la{wf}
\wfe&=&\wf-\fr{\delta}{3H\rf}= \wf-\fr{f'\dot \phi}{3Hf}\\
\wpe &=&  \wp+  \fr{\delta}{3H\rp}=
\wp+\fr{\rf}{\rp} \fr{f'\dot \phi}{3Hf}.
 \la{wp1}\eea
 For $\delta>0$ we have $\we >\wp$ and
the fluid $\rp$ will dilute faster then without the interaction
term   while $\rb$ will dilute slower since $\wfe<\wf$.

The complete evolution of $\rp$ and $\rf$ depends on the effective
equation of state parameters defined in eqs.(\ref{wf}).
Which fluid dominates at late time will depend on which
effective equation of state is smaller. The difference in
eqs.(\ref{wf}) is
\be\la{dw}
\Delta \we \equiv \wfe-\wpe=\Delta w - \Upsilon
\ee
with
\be\la{Uf}
\Upsilon=  \fr{ \dot f}{3H f}\le( \fr{\rp+\rf}{\rp}\ri).
\ee
While the sum gives the cosntraint
\be\la{sw}
\Of \wfe +\Omp\wpe= \Of w_\psi+\Omp\wp.
\ee
Clearly the relevant quantity to determine the relative growth
is given by $\Upsilon$
and if   $\Upsilon > \Delta w$
 we have  $\Delta \we <0$ and $\rb$ will dominate the universe at late times
 while for $\Upsilon < \Delta w$ we have  $\Delta \we >0$ and
 $\rp$ will prevail.

\subsection{Effective Potential $V_T$}

Form eq.(\ref{rp0}) we see that the effective (total)  potential
$V_T$ and its derivative w.r.t $\phi$ are given by
\bea\la{VT}
V_{T}&\equiv& V+ \rf=
V+\; \widetilde{\rf}\; \fr{f(\phi)}{f_o} \\
V'_{T} &=& V'+\; \rf\; \fr{f'}{f}= V'+\; \widetilde{\rf}\;
\la{dvt}\fr{f'}{f_o}
\eea
where we have   used eqs.(\ref{srf}) in the last equality
of eqs.(\ref{VT})and (\ref{dvt}). The mass of $\phi$ is given by
\be\la{m}
m^2=V_T''=V''+  \rf\; \fr{f''}{f}= V'+\; \widetilde{\rf}\;\fr{f''}{f_o}.
\ee
Using eq.(\ref{dvt}) we can rewrite
eq.(\ref{rp0}) as
\be\la{rp1}
\ddot\phi+3H\dot\phi+V_T'=0.
\ee
While the derivative of the effective potential $V_T'$
is negative the field $\phi$ will evolve to larger
values since $\dot\phi >0$. However, a
minimum of $V_T$ can be reached if $V'_T=0$  which
requires  $ f' > 0$,  since $V'<0$ by hypothesis, with
$V'=- \rf f'/f$.  Taking the time
derivative $\dot V_T'=V_T''\dot\phi+\pp_t V_T'=0$ one obtains \ci{wapp}
\be\la{dp2}
\dot\phi= -\fr{3HV'}{m^2}= \fr{3H\rf f'}{m^2 f}= \fr{3H\widetilde{\rf} f'}{m^2 f_o}
\ee
Notice  that even in the case with $V_T'=0$ we have  $\dot \phi>0$,
i.e. the $\phi$ grows with time even at the minimum
of $V_T$, and we also have   $\dot f=f'\dot\phi > 0 $.
The mass $m$ given by eq.(\ref{m}) becomes \ci{wapp}
\be\la{mm}
m^2=  V''+\rf \fr{f''}{f}= \rf\fr{f''}{f}\le(1+\fr{f'^2}{f f''}
\widetilde{\Gamma} \fr{\rf}{V}\ri)
\ee
with $\widetilde{\Gamma}\equiv V''V/V'^2$ ($\widetilde{\Gamma}_m \gtrsim 1$ if the field
$\phi$ is tracking \ci{Q}) and we used $V'=- \rf f'/f$.
Approximating $V\simeq \rp$ we get
\be\la{mH}
\fr{m^2}{H^2}\simeq 3\Of\fr{f''}{f}\le(1+\fr{f'^2}{f f''}
\widetilde{\Gamma}_m \fr{\Of}{\Omp}\ri).
\ee
Solving eq.(\ref{rp1}) with $V_T'\equiv 0$ (i.e. the
usual  slow roll condition $\ddot\phi\ll 3H\dot\phi$ is no longer
satisfied) gives $\ddot \phi+3H\dot\phi=0$ with  a solution
$\dot\phi \sim a ^{-3}$.
 Since $\widetilde{\rf}$ redshifts as $a^{-3}$
we  see from eq.(\ref{dp2}) that
\be\la{cc}
k\equiv \fr{\dot\phi}{\widetilde{\rf}}= \fr{3Hf'}{m^2 f_o}=  \fr{f}{ f_o}\fr{\sqrt{\rp(1+\wp)}}{\rf}
\ee
with $k$ a constant and we have used from eq.(\ref{wp}) $\dot\phi^2=\rp(1+\wp) $
and $\widetilde{\rf}=\rf f_o/f$. By taking
$\Omp=\rp/3H^2,\Of=\rf/3H^2$,  eq.(\ref{cc}) with $M=f$ gives
\be\la{MH}
\fr{1}{k^2} \fr{M^2}{H^2}
= \fr{3\Of^2}{\Omp\le(1+\wp\ri)}.
\ee
 Dividing
eq.(\ref{mH}) by eq.(\ref{MH}) we get
\be
\fr{m^2}{M^2}=\fr{\le(1+\wp\ri)}{k^2}\fr{\Omp}{\Of} \fr{f''}{f}\le(1+\fr{f'^2}{f f''}
\widetilde{\Gamma}_m \fr{\Of}{\Omp}\ri).
\ee
 At present time with
 $\Ompo=0.7,\Ofo=0.3,\wpo=-0.9$ we have
$\Of^2/\Omp (1+\wp )\simeq 1.3$ and $
 M^2/H^2k^2 \simeq  5$.

\section{Decay Width $\G$}\la{DeWi}

The conditions for a particle to decay is  that its lifetime
$\tau=1/\Gamma $, where $\G$ is the decay width, is smaller   than the life of the universe given
by $t\propto 1/H$,
\be\la{gh}
\fr{\Gamma}{H} > 1
\ee
and that the
energy condition (in the rest frame of the decaying particle)
\be\la{mM}
m^2  \geq \Sigma_i M^2_i
\ee
must be satisfied, where
$m$ is the mass of the decaying particle and the sum is over the
product particles with mass $M_i$.

We assume an interaction term ${\it L_{int}}= - f(\phi)
\overline{\psi}\psi$ as in eq.(\ref{int}) with $f$ not necessarily
a positive power law function of $\phi$, as for example $h\propto
e^{b\phi}$ or $h \propto 1/\phi^2$. If we expand $f$ in a Taylor
series around  a  time $t=t_e$, $ \phi_e=\phi(t_e)$, we
have  $f=f_e+f'|_e
\delta\phi +(1/2)f''|_e \delta\phi^2+...$, with
 $f'=df/d\phi$ and $f_e=f(\phi_e)$,
then the interaction term   gives an effective coupling
 \bea\la{Lint}
-{\it L_{int}}&=& f(\phi)\overline{\psi}\psi \\
&=&  f_e\overline{\psi}\psi +f'_e\,
\delta\phi\;\overline{\psi}\psi
+\fr{1}{2}f_e''\,\delta\phi^2\;\overline{\psi}\psi+...\non
\eea
between  two  fermions and  $q$ quantum scalar fields $\phi$, with
$q=1,2...$ . The first term in eq.(\ref{Lint})
gives the mass of the fermion while the second gives
an interaction between one scalar field and two fermions.
This term is responsible for the scalar decay.
At the  time $t_e$ the fermion field gets a mass
given by $M=f_e$. Since the  potential of $\phi$ might be a run
away potential, i.e. the minimum is at $\phi \rightarrow \infty $,
the expansion point $\phi_e$ is a function of time and the mass
$M$ and couplings $f_e, f'_e$ are also functions of time.

If $f=h\phi$, with $h$ constant, the decay rate of $\phi$ into two fermions
 is given  by $ \Gamma = h^2 m/ 8\pi $ and
generalizing it to an arbitrary function $f(\phi)$ we obtain
\be\la{G}
\Gamma = \fr{f'^2 m}{8\pi}.
\ee
The dark energy scalar field  will  decay
if $\G/H >1$ and $m^2>2M^2$ which implies that
\be\la{G2}
\fr{\G^2}{H^2}=   \fr{f'^4m^2}{(8\pi)^2H^2}=
\fr{3f'^4}{(8\pi)^2} \le(\fr{V''}{3H^2}+  \Of \fr{f''}{f}\ri )> 1
\ee
or
\be\la{G3}
\fr{m^2}{2M^2}=\fr{1}{2} \le(\fr{V''}{f^2}+  \rf \fr{f''}{f^3}\ri )> 1
\ee
with $m^2=V''+\rf f''/f$ as defined in eq.(\ref{m}).

\section{Generic Conditions}\la{GeCo}

The interaction between dark energy and $\rf$ must be such that it
allows  the universe to accelerate recently. Therefore the
dark energy particle must not decay before present day. However,
the dark energy may decay in the future and the universe would
then stop accelerating.
We have therefore different sets of conditions:
for $t$  smaller than $t_o$,  present day $t=t_o$  and   for times $t>t_o$.

\subsection{Conditions for acceleration}

For the universe to accelerate in the absence of any interaction
the slow roll conditions in eq.(\ref{sr}) must be satisfied.
However, once   the interaction is turned on the slow roll
conditions change.

For generality purpose we take the energy density of the universe as
$\rf+\rp$ plus $\rho_{nim}$, with $\rho_{nim}$
a non interacting matter with $w_{nim}=0$.
The Hubble parameter is   then
\bea
3H^2 &=& \rp+\rf+\rho_{nim}=\fr{1}{2}\dot\phi^2+V(\phi)+\widetilde{\rf}\;
\fr{f(\phi)}{f_o} +\rho_{nim}\non\\
&=& \fr{1}{2}\dot\phi^2+V_T +\rho_{nim}.
\eea
Using eqs.(\ref{rmp}) and (\ref{rmf}) it is easy to
see that the universe accelerates if
\bea\la{acel}
\fr{\ddot a}{a}&=& H^2+\dot H = -\fr{1}{6}\le(\rho_m+\rf+\rp+3p_\phi \ri)\non\\
&=& -\fr{1}{6}\le(\rho_m+2\dot\phi^2 + V_T - 3V\ri)\\
&=&  -\fr{1}{6}\le(\rho_m+2\dot\phi^2 + \rf -2 V \ri)\non
\eea
is positive.
A positive
acceleration gives the  constraint
\be\la{cv}
\fr{1}{2}\fr{\rho_{nim}+\rf+\dot\phi^2}{V} < 1.
\ee
Clearly $V$ must
be larger than $(\rho_m+\rf)/2$ and  $\rp>(\rho_m+  \rf)/2$.
Eq.(\ref{cv}) gives the usual constraint $ \Omp\wp<-1/3$,
even though $\rf$ is also a function of $\phi$.
So as long as $\Omp< 1/3$, or
$\Omega_{nim}+\Of >2/3$,   there will be no acceleration.
For recent times  $\rp>\rf$ but well before
that we will have the dark energy as subdominant
energy density  $\rp \ll  \rho_m\equiv\rho_{nim}+\rf$.

We would like to emphasize that the "naive" conditions
on the total scalar potential $V_T $
\be\la{slvt}
  |\fr{V'_T}{V_T}|, \hspace{1cm} |\fr{V''_T}{V_T}|<1
\ee
{\it do  not} imply an accelerating epoch. The
reason is that only the potential $V$ and not
$V_T$ enters with a negative sign (within the brackets)
in eq.(\ref{acel}). The proof is simple, just take
the limiting case $V=V'=V''=0$, in this case
$V_T=\rf$ and $  V'_T/V_T=f'/f, V''_T/V_T=f''/f$
can be both much smaller than one but
condition in eq.(\ref{cv}) is not satisfied
and therefore the universe does not accelerate.

\subsection{Conditions for $t<t_0$}

For dark energy not to decay before present day we need
the constraints in eqs.(\ref{G2}) and (\ref{G3}), $\Gamma/H<1$ or $m^2<2M^2$,
not to be satisfied simultaneously,    therefore we need
\be\la{c1}
\fr{\G^2}{H^2}=   \fr{f'^4m^2}{(8\pi)^2H^2}=
\fr{3f'^4}{(8\pi)^2}  \le(\fr{V''}{3H^2}+  \Of \fr{f''}{f}\ri )< 1
\ee
or
\be\la{c2}
\fr{m^2}{2M^2}= \fr{1}{2} \le(\fr{V''}{f^2}+  \rf \fr{f''}{f^3}\ri )< 1
\ee
Clearly the interaction term $f(\phi)$ and its derivatives
play the crucial  roll. The  term $V''/3H^2 $ in eq.(\ref{c1}) is much
smaller than one
by hypothesis so the inequality requires
 $3f'' f'^4/f(8\pi)^2<1/\Of$.

\subsection{Present day $t=t_0$}

At present day we know that the universe is dominated by dark energy
with
\be
\Ompo=0.73, \hspace{1.5cm} \Omega_{m o}=0.27
\ee
and dark matter $\Omega_m$ includes  non interacting dark matter
$\rho{_nim}$   and interacting dark matter $\rho_{im}$,
i.e. $\rho_m=\rho_{nim} +\rho_{im}$. In our case the interacting
dark matter is given by fermions, $\r_{im}=\rf$. For
simplicity we could take $\rho_m=\rf,\;\rho_{nim}=0$.

The cosmological data for a constant
 equation of state of a dark energy, assuming no interacting
with other fluids, is \ci[DE].
\be
<w_{ap}>=-1.04\pm 0.06.
\ee
The equation of state of sate   $\wp$  (given by
eq.(\ref{wp})),  is  $\wp\geq -1$ but an apparent $\wa<-1$,
as defined in eq.(\ref{wap}),
can be obtained due to the interaction
between $\phi$ and the fermion fields. Since
$\wa$ is smaller than -1 we take this as an indeication
od dark energy interaction.

At present time $\rp>\rf$ and using
the condition for acceleration in eq.(\ref{cv}) we need
\be
 \fr{\rf} {V}|_o<2
\ee
where we have taken for simplicity $\rho_{nim}=0$ and
the slow roll conditions in eq.(\ref{sr}), which imply
that $\dot\phi^2/V\ll 1$.

\subsection{Conditions for  dark energy decay for $t>t_o$}

Dark energy will stop dominating the universe
  if either
$\rf$ redshifts slower than $\rp$, due to
the interaction terms $f(\phi)$, or  if the dark energy
particle decays into the fermion fields.

Dark energy  density will dilute compared to the
fermion energy density if
\be
\Upsilon=  \fr{ \dot f}{3H f}\le( \fr{\rp+\rf}{\rp}\ri)>\Delta w .
\ee
Dark energy decay will take place if
$\Gamma/H > 1$   and  $ m^2 > 2M^2$ given
by eqs.(\ref{G2}) and (\ref{G3}).

\section{Fluid Evolution}\la{FlEv}

The complete evolution of $\rp$ and $\rf$ depends on the effective
equation of state parameters defined in eqs.(\ref{wf}).
  The difference in
eqs.(\ref{wf}) is \ci{fate}
\be
\Delta \we \equiv \wfe-\wpe=\Delta w - \Upsilon
\ee
with $\Delta w \equiv w_\psi-\wp$ and
\be\la{u}
\Upsilon\equiv \fr{\delta}{3H}\le(\fr{\rp +\rb}{\rp\rb}\ri).
\ee
If we take the ratio $y\equiv \rp/\rb=\Omp/\Of$
 the derivative of $y$ w.r.t. time is \ci{fate}
\be\la{yt}
\dot y= 3Hy\le[\Delta w  - \Upsilon\ri].
\ee

The value for $y$ is constraint to
$0\leq y\leq \infty$ with $y=0$ for $\rp=0$ and
$y=\infty$ for $\rb=0$.
Clearly from eq.(\ref{yt}) we see that
the evolution of $y$ depends on the sign of $\Delta w - \Upsilon$.
Clearly the relevant quantity to determine the relative growth
is given by $\Upsilon$
and if   $\Upsilon > \Delta w$
 we have  $\Delta \we <0$ and $\rb$ will dominate the universe at late times
 while for $\Upsilon < \Delta w$ we have  $\Delta \we >0$ and
 $\rp$ will prevail. For no interaction $\delta=0$  and  $\Upsilon=0$ gives
$\Delta \we=\Delta w>0$ and $\rp$ dominates at late times.
If $\Upsilon = \Delta w$ then $\wfe=\wpe $ and
 the ratio of both fluids $\rf/\rp$ will approach a constant value, and
if the universe is dominated by
$\rp+\rf$, i.e. $\Omp+\Of=1$, then eq.(\ref{sw}) gives
\be
\wp\leq\; \we = \wf\Of+\wp(1-\Of) \;\leq \wf,
\ee
i.e. the effective equation of state is constraint between $\wp$ and $\wf$.

\subsubsection{Non Interaction solution: $\Delta w > \Upsilon $}\la{ni}

If $(\wf-\wp)> \Upsilon $ then $\dot y$ is positive and
$y$ will increase, i.e. $\rb$ will dilute  faster than $\rp$,
and we will end up with $\rp$ dominating the universe. In this
case the interaction term $\delta$ is subdominant and the evolution
of $\rp$ and $\rb$ is the usual one, i.e.
$\rp \propto a^{-3(1+\wp)}$ and $\rb\propto a^{-3(1+\wf)}$.

\subsubsection{Finite solution: $\Delta w=\Upsilon $}\la{=}

A  solution to eq.(\ref{yt}) with $y$ constant and
$\rb\neq 0, \rp\neq 0$ is only possible if $\delta/H\rp$
is positive and constant since
$\Upsilon=\delta(1+y)/H\rp=\Delta w$ must
be constant and positive,
taking $\Delta w=w_\psi-\wp>0$ constant. Let us take
  $\delta= C H \rp$ with $C$ a positive constant,
and from eq (\ref{yt}) for $\dot y=0$
we get a stable value of $y$ given by \ci{fate}
\be\la{ys}
y_s =\fr{\Omp}{\Of} =\fr{3(\wf-\wp)}{C}-1.
\ee
It is easy to see that the solution $y_s$ is stable
since from eq.(\ref{yt}) the fluctuation $\delta y= y-y_s$ to first order
behaves as $\delta \dot y/\delta y =-CH y_s $ ($CH$ is
positive by hypothesis) giving $\delta y\rightarrow 0$.
The evolution of $\rp$ is given by \ci{fate}
\be\la{rp3}
\rp =\rpi \le(\fr{a}{a_i}\ri)^{-3(1+\wp)-C}=
\rpi \le(\fr{a}{a_i}\ri)^{- 3(1+\fr{\wf+y_s\wp}{1+y_s})}
\ee
where we have used that $C=3(\wf-\wp)/(1+y_s)$.
Since $C$ is positive the solution in eq.(\ref{rp3})
dilutes faster than the non interacting solution
$\rp \propto a^{-3(1+\wp)}$ and with an effective equation of state
$\wpe=\wp+C/3=(\wf+y_s\wp)/(1+y_s)>\wp$.

\subsubsection{Interacting solution: $\Delta w < \Upsilon $}\la{in}

For $\rb$ to dominate the universe we need  $y\ll 1$ at late time
and the (interaction) term $\Upsilon$ should dominate over $\Delta w$.
A simple example is when $A=\delta/\rp$ is constant and positive. In this case
the evolution of $y$ is given by
\be
y=y_i\le(\fr{a}{a_i}\ri)^{3(\wf-\wp)}\,e^{-A t} \rightarrow 0
\ee
at late times for any value of $\wf-\wp$.
Now,
let us take $A$ as a constant decay rate $\G$.\\
\noindent
I) Constant decay Rate $\G=A=\delta/\rp$\\

If the mean lifetime of a particle
is given by a constant $\tau$ then we can expect it to
decay into lighter fields. In the case
 $\G=1/\tau$ constant
the constrain $\G>3H$  will  be satisfied
at some time $t\sim \tau$ (or equivalently at
$\G/H \sim 1$).

For a constant $\G$ the  interaction term is $\delta=\G\rp$
and the solution  to  eq.(\ref{rmf}) using (\ref{wp1}) is simply
\be\la{rp2}
\rp = \rpi \le(\fr{a}{a_i}\ri)^{-3(1+\wp)}\,e^{-  \G t}.
\ee
For small $t$, i.e. for $3H(1+\wp) \gg \G $,
one has the usual redshift in the absence of any interaction
$\rp\propto a^{-3(1+\wp)}\propto 1/t^2$
while for large $t$, i.e. $3H(1+\wp) \ll \G $, one  has an exponential decrease
 $\rp\propto e^{-t \G}$.
Since $\G$ is constant,  then clearly $ \G \rp/\rb$ will
 not be constant   and the solution to eq.(\ref{rmf}) will be quite
 different to that of $\rp$.

 Let as take
 the ansatz $\rb=\rbi a^{-3(1+\wf)}+g(t)\rp$ taking the time
 derivative we have
 \bea\la{rbt}
 \dot\rb &=&-3H(1+\wf) \rb + \dot g \rp +g \dot\rp \\
 &=& -3H(1+\wf) \rb + \le(3H\Delta w g -\G g+ \dot g  \ri)\rp\non
\eea
 and eq.(\ref{rbt}) should be  equal to eq.(\ref{rmp}), i.e. $
\G=\delta/\rp=3H\Delta w g + \dot g -\G g$. Therefore, the
function $g$ must satisfy the equation
\be\la{f}
 \G=  \fr{3\Delta w H g  + \dot g}{1+g}.
\ee
Eq.(\ref{f}) is valid for any functional form of $ \G$ and not
necessarily a constant. However, a simple solution can be found
when $\G$ is constant for small $t$. The solution is
$g=g_i t$ giving $\G\simeq (3\Delta wH_i  + 1)g_i$
constant,
and using for the dominant energy density $H_i=2/3(1+\wp)$
we have $\G=g_i (1+2\wf-\wp)/(1+\wp)$. The solution to eq.(\ref{rbt}) is then
\be\la{rb2}
\rb=\rbi a^{-3(1+\wf)}+ q\; \rp \G\,t
\ee
with $q= (1+\wp)/(1+2\wf-\wp)$ (if we have matter decaying
into radiation, $\wp=0$ and $\wf=1/3$, then $q=3/5$).
Eq.(\ref{rb2}) shows that the fluid redshifts as usual for $ \G t \ll 1$,
 i.e. $\rb \sim a^{-3(1+\wf)}$,
where the first term in eq.(\ref{rb2})
dominates. At
 $t\simeq t_i(\G t_i)^{-q}$ the second term starts to dominate
giving $\rb\propto \rp t\propto t^{-1}\propto a^{-3(1+\wp)/2}$,
with the fluid $\rp$ decaying  into  $\rb$ at around $\G t \simeq 1$
when   $\rp$ starts decreasing exponentially fast. For
$\G t\gg 1$ the fluid $\rb$ will dominate the universe redshifting
again as  $\rb \sim a^{-3(1+\wf)}$.

We have seen that a constant decay rate $\G$ gives an exponentially
suppressed energy density $\rp$. However, a constant decay
rate is not realistic. The decay rate $\G$ for a scalar
field decaying into fermions, given in eq.(\ref{G}), is a function
of $m$ and $f'$ and  both are in general field and time dependent
so $\G$ will   not be constant.

\subsection{ Apparent Equation of State}\la{ApEq}

 An interesting result of the interaction between   dark
 energy "DE"  with other particles
 is to change  the apparent
 equation of state of dark energy
 \ci{IDE}-\ci{wapp}. An observer that supposes
 that DE has no interaction sees a different
 evolution of DE as an observer that takes into account
for the interaction between DE and another fluid. This effect allows
to have an  apparent equation of state $w<-1$ for the
``non-interaction" DE \ci{wapp}  even though the true equation of
state of  DE is larger than -1.

Let as take the
 energy density   $\rho=\rp+\rb=\rho_{DE}+\widetilde{\rb}$.
 The energy densities $\rp,\rb$ are given by eqs.(\ref{rmf})
 and (\ref{rmp}) and these two fluid interact via  the
 $\delta $ term. On the other hand the energy
 densities $\rho_{DE}$ and $\widetilde{\rb}$ do not interact with each other
 by hypothesis and  therefore we have
$\dot{\widetilde{\rb}}=-3H$ (we have taken $\wf=0$)  and
$\dot\rho_{DE}=-3H\rho_{DE}(1+w_{ap})$, i.e.
\bea
\widetilde{\rb}&=&\rbo a^{-3 }\non\\
\rho_{DE}&=&\rho_{DE o} a^{-3(1+w_{ap})}
\eea
 if $ \wa$ is constant.
It was pointed out  that the apparent equation of state $w_{app}$
can take values smaller than -1 and it is given by \ci{wapp}
\bea\la{wap}
  w_{ap}&=&\fr{\wp}{1-x}\\
x\equiv -\fr{\widetilde{\rb}}{\rp}\le(\fr{f}{f_o}-1\ri)
&=& -\fr{\rf}{\rp}\le(1-\fr{f_o}{f}\ri)\non
\eea
  and we have used eq.(\ref{srf}).
We see from eq.(\ref{wap})
that for $\rb<\widetilde{\rb}$, i.e.  $f(t)<f_o(t_o)$
for $t<t_o$,
we have $x>0$ and $w_{ap} < \wp$
which allows to have a $w_{ap}$ smaller than -1 in the
past. However,  for $f>f_o$ with $x<0$ and $w_{ap}> \wp$ for $t>t_o$,
The observational evidence shows
that $f$ is at present time  growing function of $\phi$, i.e. $f'>0$.

\subsection{Evidence for Dark Energy Decay }\la{ded}

The cosmological observations prefer an equation of state $w<-1$ for dark energy.
In principal a $w<-1$ for a fluid is troublesome since it has
instabilities and causality problems. However, as seen in section
\ref{ApEq} this  can be an optical effect due to the interaction between
dark energy with other particles as for example   fermions.
It was shown that  for  an  effective
$w<-1$,  interpreted as an  interaction between dark energy
and another fluid,
can be a signal for  dark energy decay in the future with
the universe  no longer accelerating \ci{fate}.

Lets us expand $f$ around present time $t_o$
$f=f_o+\dot f_o \delta t$ and  we
keep only the first order term. Now, from eq.(\ref{wap})
 $x$ can be written as \ci{fate}
\be\la{x2}
x =-\fr{\widetilde{\rf}}{\rp}\le(\fr{f}{f_o}-1  \ri)
\simeq -\fr{\widetilde{\rf}}{\rp} \fr{\dot f_o}{f_o} \delta t=
-\fr{\rf}{\rp} \fr{\dot f_o}{f}  \delta t
\ee
with $\widetilde{\rf}=\rf f_o/f,\;\delta t\equiv t-t_o$.
For $t<t_o$ we have $x>0$ since $\dot f_o>0$.
The interaction term $\delta =\rf \dot f/f$, is now
\ci{fate}
\be\la{del1}
\delta =-\rf\fr{x}{\delta t}
\ee
where we have taken $\dot f=\dot f_o$, consistent
with the approximation taken for $f$.
Let us take the simple phenomenological ansatz   for $x$
given as   $x=\beta (a_o-a)=-\beta \delta a$ with $\beta$
a constant to be determined by observations.
The SN1a data are
in the range $1>a>2/5$, i.e. for a redshift
$0<z<1.5$, and the best fit solution has  an average equation of
state $<w>\approx -1.1$ \ci{DE}. Taking the average of
$w_{app}=\wp/(1-x)$ we have
 \be\la{wa2}
 <\wa>\equiv
\fr{\int_{a_1}^{a_o} \wa\,da}{\int_{a_1}^{a_o}da}= \fr{\wp
Log\le[1-\beta(1- a_1)\ri]}{\beta (a_1-1)}
\ee
where we have used
$x= -\beta \delta a$  and $a_o=1$. As an example let as take $<\wa>=-1.1$,
as suggested by the observations \ci{DE,SN},\ci{IDE-ax}, and
$ \wp=-0.9,w_\psi=0, \;a_1=2/5 $.
In this case we obtained from eq.(\ref{wa2}) the value $\beta=0.56$
\ci{fate}.
Using the anstaz $x= -\beta \delta a$
  with $ \delta a=a H\delta t$ then eq.(\ref{del1}) becomes \ci{fate}
\be\la{d}
\delta= \beta a H \rf, \hspace{.2cm} \fr{\dot f}{f}= \beta aH
\hspace{.2cm} and \hspace{.2cm} \Upsilon =\fr{\beta a}{3}\le(1+\fr{\rf}{\rp}\ri).
\ee
Using eq.(\ref{rmf})   with the interaction
term given in eq.(\ref{d}) we get an energy density \ci{fate}
\be
\rp=\rpo \, a^{-3(1+\wp)}e^{-\beta(a-1)}
\ee
which shows that $\rp$  dilutes as $a^{-3(1+\wp)}$ for $a\ll a_o=1$
and $\rp$ is exponentially suppressed for $a \gg 1$.
In this case $3\Delta w < \Upsilon$ and $\Upsilon \rightarrow\infty$ at late times,
as seen from eq.(\ref{d}).

\section{Dark Energy Disappearance }\la{dedis}

The decay rate $\G=f'm^2/8\pi$, the mass $m$ and
$M$  are a field and time dependent quantities.
If we take $f$ as a monotonic function, since
$f'$ at present time is positive (cf. section \ref{ded}), then $f$ should
be an increasing function of $\phi$ for all
values of $\phi$. This implies that the mass of
the fermion $M=f$ increases at all times. So,
if the mass of the scaler field $m$ vanishes
asymptotically then $m/M$ will be smaller than one
at some point in the future but larger in the past.
So, we would expect on general grounds that
the condition $m^2>2M^2$ would be easily
satisfied in the past. Therefore we require
$\G/H<1$ and from eq.(\ref{c1}) we need
\be
\fr{3f'' f'^4}{(8\pi)^2 f}<1
\ee
where  we have taken $\rf\gg \rp$ ($\Of=1$) in the far past.

\subsection{Dark Energy Dilution}\la{dedil}

As we have seen in section \ref{FlEv} dark energy
dilution, $\rp\gg \rf$ at late times, will only take place if $\Upsilon > \Delta w$ with
\be\la{Uf2}
\Upsilon=  \fr{ \dot f}{3H f}\le( \fr{\rp+\rf}{\rp}\ri).
\ee
Let us take  $\dot f/fH=(a/f)(df/da)$ with the ansatz
$f=f_o a^q$.  We then have $\dot f/fH=q $,
$\Upsilon =q/3$ and the interaction term
$\delta= q\rf H$. For $q>q_o\equiv 3\Delta w$ we
have $\wfe< \wpe$ and $\rf$ will dominate while for
$q<q_o$ then  $\rp$ will end up dominating the universe.
Furthermore, if $f$ grows faster than power law then  $\Upsilon
\rightarrow\infty$ and $\rf$ will dominate at late times. this
is the case discussed in section \ref{ded} as sen from eq.(\ref{d}).
For
$f$ growing slower than power law we have $\Upsilon \rightarrow 0$
and $\rp$ will end up dominating the universe.

The limiting case is for $ \rf/\rp$  constant
and a solution  is obtained for $\Upsilon = q/3=\Delta w$.
This case is equivalent as in section \ref{=}
with $\rf \leftrightarrow \rp$ and $q=C$.
From eq.(\ref{ys}) we have
\be
\fr{\rf} {\rp}=\fr{\Omega_\psi}{\Omp}=\fr{3\Delta w}{q}-1
\ee
(see discusion belown eq.(\ref{ys})).

Now,  we consider again the interaction term $f(\phi)$ as a
function of a scalar field $\phi$ with $\dot f=f' \dot\phi $.
The evolution of $\phi$ is determined by eq.(\ref{rf}) and the
sign of $V_T'$, defined in eq.(\ref{VT}),  sets if $\phi$ grows or
decreases with time. We will discuss the cases $V'_T>0$,
$V'_T< 0$ and $V'_T = 0$ separately.

\subsubsection{Case:  Positive Derivative $V'_T>0$}

First, let us consider the possibility
that $V_T'> 0$ which implies from eq.(\ref{rp1}) that $\dot \phi<0$.
In this case since $V'<0$ then $f'>0$
and  $\dot f= f'\dot\phi$ is negative giving a negative
$\Upsilon$ (c.f. eq.(\ref{Uf2})) and $\rp$ will dilute slower
than $\rf$ (even slower than without the interaction).

\subsubsection{Case: Negative Derivative  $V'_T<0$}

For $V_T'<0$
we have  $|V'|> \rf f'/f$ and  from eq.(\ref{rp0}) we can
approximate
\be\la{dp3}
\dot\phi\simeq -V'/3H.
\ee
Taking, for simplicity,
$\rp+\rf=3H^2$ in eq.(\ref{Uf2}) we get
\be\la{ff}
\Upsilon= \fr{H \dot \phi f' }{\rp f} = -\fr{f'}{f}\fr{V'}{3\rp} >\Delta w.
\ee
The condition
$V_T'<0$ implies $f'/f<|V'|/\rf$  and together with the slow roll
conditions  $\rp>V>|V'|$ and  eq.(\ref{ff}) we get  the constraint
 \be\la{co1}
 \fr{\Omp}{\Omega_\psi}>\fr{f'}{f}>3\Delta w
\ee
which sets an upper value
\be\la{co2}
\Omega_\psi < \fr{1}{1+f'/f} <
\fr{1}{1+3\Delta w}.
\ee
If we take for example $\wf=0$ and
$\wp\approx -1$ we have $\Delta w=1$ and $\Of < 1/4$.
Therefore, we conclude that in the case $V_T<0$
it is not possible for $\rf$ to dominate the universe.

\subsubsection{Case:  $V'_T=0$}

As we have seen, while the derivative of the effective potential $V_T'$
is negative the field $\phi$   evolves to larger
values since $\dot\phi >0$. However, a
minimum of $V_T$ can be reached if $V'_T=0$, which
requires  $ f' > 0$  since $V'<0$ by hypothesis, and
$V'=- \rf f'/f$.  Taking the time
derivative $\dot V_T'=0$ one obtains
with
\be\la{dp4}
\dot\phi= -\fr{3HV'}{m^2}=  \fr{3H \rf   f'}{m^2 f_o}
\ee
(c.f. eq.(\ref{dp2})) which is still positive, i.e.
even tough $V_T'=0$ the field $\phi$ still grows
with time. The reason  being that $\widetilde{\rf}$
is a function of time (through $a(t)$).
Calculating $\Upsilon= H \dot \phi f'/\rp f > \Delta w$
and using eq.(\ref{dp4}) and $1> |V'|/V= \rf f'/fV$ we get the constraint

\be\la{co3}
\fr{\Omp}{\Omega_\psi}>\fr{f'}{f}>\fr{m^2}{3H^2}\Delta w
\ee
which sets an upper value
\be\la{co4}
\Omega_\psi <
\fr{1}{1+(m^2/3H^2)\Delta w}.
 \ee
Comparing $\dot\phi$ form
eqs.(\ref{dp3}) (labeled $\dot\phi_1$) and (\ref{dp4})
(labeled $\dot\phi_2$) we see that the
ratio $\dot\phi_1/\dot\phi_2=m^2/9H^2$
is exactly that in the denominator of   eqs.(\ref{co2}) and
(\ref{co4}). For $\dot\phi_1/\dot\phi_2=m^2/9H^2\ll 1$
the evolution of $\phi$ is faster in the case $V_T=0$
than for $V_T<0$. Now, if $m^2/3H^2\geq 1$ and taking
$\Delta w=1$ we have $\Omega_\psi\leq 1/2$.
However, if $m^2/3H^2\ll 1$ than then the constraint in eq.(\ref{co4})
becomes $\Of \approx 1$ allowing for the fermion
fields to dominate at late times.

It is easy to see from eq.(\ref{co3}) that in the limit $\Omp \rightarrow 0$
we need $f'/f\rightarrow 0 $ and $ m^2/3H^2 \rightarrow 0$. From eq.(\ref{mm})
the condition
$m^2/3H^2 \rightarrow 0$ implies that $ f''/f\rightarrow 0$ since $\Of=1-\Omp=1$  and
$V''/V\ll 1$ by hypothesis.

\subsection{Dark Energy Decay}\la{dedec}

The conditions for dark energy to decay into fermions
given in eqs.(\ref{G2})
and (\ref{G3}) are constraints on the dark energy mass $
m^2= V''+\rf f''/f$.
Let us consider the  two  possible  cases $V''\geq \rf |f''/f|$
and $V''\leq \rf |f''/f|$ separately.

\subsubsection{Case: $m^2\simeq V''$, i.e. $V''\geq \rf |f''/f|$.}

Using the slow roll hypothesis on $V$ given in eqs.(\ref{sr})
  we can set
the upper limit
$\rp> V>V''>\rf |f''/f|$ which gives the constraint
\be\la{de1}
\fr{\Omp}{\Of}>|\fr{f''}{f}|.
\ee
If we take $\Omp+\Of=1$ then eq.(\ref{de1}) gives un upper limit
$\Of<1/(1+|f''/f|)$ and in the case $\Omp\rightarrow 0, \Of \rightarrow 1$ we require
$|f''/f|\rightarrow 0$.

The constrain in eq.(\ref{G2}) gives
$V''>H^2/f'^4(8\pi)^2$ and using $H^2>V/3$ and $V>V''$ we obtain
\be\la{de2}
f'^2>\fr{1}{24\pi}
\ee
while for condition in eq. (\ref{G3}) we have
\be\la{de3}
1>2\fr{f^2}{V''}\geq 0.
\ee
Since $V'' \rightarrow 0$ by hypothesis,  eq.(\ref{de3}) implies that $f \rightarrow 0$
at late times, i.e. $f$ is a decreasing function at large $t$,
since $M=f\geq 0 $. Eq.(\ref{de3})  can only be satisfied if $f'$ is negative,
i.e. $f$ is a decreasing function. However,
at present time  the dark energy interacting
must have $f'_o>0 $ if $w<-1$, as seen in section (\ref{ded}).
So, unless $f$ increases at present time and than it starts decreasing
at a later stage (i.e. it is no longer  a monotonic function of $\phi$)
the conditions for a dark energy decay into fermions with $V''>\rf f''/f$ cannot take
place. However,  $f$ will  no longer be  a monotonic function of $\phi$.

\subsubsection{Case: $m^2\simeq \rf f''/f$, i.e. $ \rf f''/f \geq V''$.}

Let us now take the case $m^2\simeq \rf f''/f> V'' $, which  implies that
\be\la{de4}
0\leq \fr{V''}{f''}<\fr{\rf}{f}=\fr{\widetilde{\rf}}{f_o}\propto a^{-3}
\ee
where we used eq.(\ref{srf}). Eq.(\ref{de4})  shows that at late times
(with $a\rightarrow \infty$) we have  $V''/f''\rightarrow 0$,
i.e. $V''$ must decrease faster than $f''$.
Using $m^2\simeq \rf f''/f$ and condition  in eq.(\ref{G3}) we get in this case
the constraint
\be\la{de5}
0\leq \fr{f}{f''} < \fr{\rf}{2f}= \fr{\widetilde{\rf}}{2f_o}\propto a^{-3}
\ee
where we have used again eq.(\ref{srf}). Once again we see that at
$a\rightarrow \infty$ that  $f/f'' \rightarrow 0$ at late times.
Finally, the constraint in eq.(\ref{G2}), with $m^2\simeq \rf f''/f$
and  $3H^2>\rf$, gives
\be\la{de6}
\fr{f''f'^4  }{f} > \fr{1}{3(8\pi)^2}  .
\ee
Since from eq.(\ref{de5}) $f/f''$ must vanish at late times,
eq.(\ref{de6}) is satisfied as long as
$f'^4$ does not go to zero  faster
than $f/f''$.

\subsection{Examples}

To have dark energy decaying  the conditions
$V''/f'' \rightarrow 0$,  $ f/f''\rightarrow 0$ and $f''f'^4/f>1/3(8\pi)^2$
must be satisfied. A simple example is
\be
f=f_oe^{\alpha \phi^2}
\ee
 with $\alpha $ a positive constant. Since
$\phi$ grows with time, so does $f, f',f''$ and the constraint
$V''/f''\ll 1 $ and  $f''f'^4/f > f^4>1/3(8\pi)^2$ are
satisfied trivially and the condition
\be
\fr{f''}{f}=2\alpha +4 \alpha^2\phi^2 \gg 1
\ee
is also satisfied for $\phi\rightarrow \infty$.
Clearly, this potential will not work
for dark energy dilution.

 Dark energy diluting needs
$f'/f \rightarrow 0 $ and $ f''/f\rightarrow 0$ and
an example is
\be
f=f_o \phi^\alpha
\ee
with $\alpha\geq 2$. In this case
$f'/f\sim 1/\phi$ and $f''/f\sim 1/\phi^2$
and $\phi$ grows with time.

\subsection{Summary}

To conclude, we have seen that if the conditions
on the interaction term
$f'/f\rightarrow 0 $ and $ f''/f\rightarrow 0$ are satisfied
then we can have a dark energy diluting faster
than the energy density of fermions. In this
case we will have $m^2/H^2\ll 1$.
On the other hand,  for  dark energy  to
 decay into fermion fields we require
that dominant part  of the scalar mass is
given  by the interaction term $m^2\simeq  \rf f'' / f \gg V'' $.
In this case, the conditions to be satisfied
are given by eqs.(\ref{de4}), (\ref{de5}) and (\ref{de6}),
i.e.$V''/f'' \rightarrow 0$,  $ f/f'' \rightarrow 0$ and $f''f'^4/f>1/3(8\pi)^2$.

The condition for dark energy redshifting faster than fermions
$ f''/f\ll 1$ and the condition for dark energy decay into
fermions $ f/f'' \ll 1$ cannot be simultaneously meet.
Clearly form eq.(\ref{cv}) if the universe is dominated by $\rf$
than it will not accelerate. We have also seen simple
examples of the interaction term $f$ which require no
fine tuning of any parameter.

\section{Conclusions}\la{conc}

A dark energy component is responsible for the present stage
of acceleration of our universe. If no fine tuning is assumed
on the dark energy potential it is easy to see that since
it dilutes slower than the other fluids, e.g. matter, then
it will end up dominating the universe at late times
and the universe will not stop this stage of acceleration
anymore.

In this {\it letter} we have studied the possibility that
the universe will stop accelerating and that dark energy
decays into fermion fields. The interaction
between  dark energy and fermions is given through
a fields dependent mass  $M=f(\phi)$ of the fermion fields.

Furthermore, the fact that the equation of state of dark energy
seems to be smaller than minus one as suggested by the cosmological data
can be an indication that dark energy does indeed interact
with other fluids. We take this fluid to be fermions.
Using the observational data we can
determine  present day value of  the interacting function $f(\phi)$
and its derivative and we can then  extrapolate the
result into the future.

The interaction term $f(\phi)$ plays an important roll in the
evolution of the dark energy. We determine the conditions under
which a dark energy can dilute faster than the fermion fields
or it can decay into these fields. We obtained that only in the
case where the mass of the dark energy, given by
$m^2=  V''+\rf f''/f$, is dominated by the interaction
term can the dark energy decay. While for dark energy
diluting faster then the fermion fields   the conditions
$f'/f\\rightarrow 0$ and $f''/f\rightarrow 0$ must be satisfied.

We have shown that naive slow roll conditions on the
effective potential $V_T$ do not imply an accelerating epoch.
The condition needed is that the scalar potential $V$ dominates
the universe (c.f. eq.(\ref{cv}))

We have seen, therefore,  that it is indeed possible to
live now in an accelerating epoch dominated by the dark
energy and  without introducing any fine tuning parameters
the dark energy can either dilute faster or decaying into fermions
in the future. The acceleration of the universe
will then cease.

\begin{acknowledgments}

This work was also supported in
part by CONACYT project 45178-F and DGAPA, UNAM project
IN114903-3.

\end{acknowledgments}

\end{document}